\documentclass[aps,reprint,twocolumns,pre,superscriptaddress,floatfix,notitlepage,nofootinbib]{revtex4-1}
\usepackage{amssymb,amsmath,amsfonts} 
\usepackage{mathtools}
\usepackage{physics}
\usepackage{array}
\usepackage{graphicx,epsfig,xcolor}
\usepackage[colorlinks=true,citecolor=blue,linkcolor=red]{hyperref}
\usepackage{newtxtext,newtxmath}

\begin{document}

\title{Chaos, thermalization and breakdown of quantum-classical correspondence in a collective many-body system}

\author{\'{A}ngel L. Corps}
    \email[]{corps.angel.l@gmail.com}
    \affiliation{Institute of Particle and Nuclear Physics, Faculty of Mathematics and Physics, Charles University, V Hole\v{s}ovi\v{c}k\'{a}ch 2, 180 00 Prague, Czech Republic}
    \affiliation{Grupo Interdisciplinar de Sistemas Complejos (GISC),
Universidad Complutense de Madrid, Avenida Complutense s/n, E-28040 Madrid, Spain}
    
    \author{Sebastián Gómez}
    \affiliation{Instituto de Estructura de la Materia, IEM-CSIC, Serrano 123, E-28006 Madrid, Spain}

\author{Pavel Stránský}
    \affiliation{Institute of Particle and Nuclear Physics, Faculty of Mathematics and Physics, Charles University, V Hole\v{s}ovi\v{c}k\'{a}ch 2, 180 00 Prague, Czech Republic}
    
\author{Armando Rela\~{n}o}
        \affiliation{Grupo Interdisciplinar de Sistemas Complejos (GISC),
Universidad Complutense de Madrid, Avenida Complutense s/n, E-28040 Madrid, Spain}
\affiliation{Departamento de Estructura de la Materia, F\'{i}sica T\'{e}rmica y Electr\'{o}nica, Universidad Complutense de Madrid, Avenida Complutense s/n, E-28040 Madrid, Spain}

\author{Pavel Cejnar}
    \affiliation{Institute of Particle and Nuclear Physics, Faculty of Mathematics and Physics, Charles University, V Hole\v{s}ovi\v{c}k\'{a}ch 2, 180 00 Prague, Czech Republic}

\date{\today} 

\begin{abstract}
We investigate thermalization and the quantum-classical correspondence in the collective Bose–Hubbard model, focusing on the four-site case. Our analysis of the classical phase-space structure and its excited-state quantum phase transitions leads us to three dynamical regimes: symmetry-breaking low-energy states, an intermediate region where quantum and classical equilibrium states markedly disagree, and a high-energy regime with restored correspondence. The observed classical intermittency above the first excited-state quantum phase transition contrasts with quantum dynamics, which remains trapped in symmetry-breaking sectors despite the existence of a classically connected phase. This mismatch originates from the population of imbalance-carrying eigenstates and persists even for relatively large number of particles. Our results reveal unexpectedly slow convergence to the classical limit, signaling robust finite-size effects in collective many-body dynamics.
\end{abstract}

\maketitle

\section{Introduction} 

The thermalization of isolated quantum many-body systems has attracted sustained interest in recent years. 
For generic non-integrable systems with short-range interactions, thermalization is commonly understood within 
the framework of the eigenstate thermalization hypothesis (ETH) \cite{Srednicki1994,Srednicki1999,Alessio2016}, which asserts that expectation values of local 
observables in individual many-body eigenstates coincide with those predicted by statistical ensembles. At its core, the ETH is a theory for ergodic behavior beyond the random matrix theory (RMT) \cite{Mehta2004,Bohigas1984,Bohigas1991,Kos2018}, which provides the fundamental apparatus for the statistical description of eigenlevels in chaotic quantum systems. This interest has been stimulated by fascinating fundamental questions as well as by potential applications that are today within experimental reach \cite{Rigol2008,Kaufman2016,Neill2016,Evrard2021}. 
When the ETH holds, unitary time evolution leads to effective equilibrium states with thermal properties that depend only on 
conserved quantities and are largely independent of microscopic details of the initial state. 
Substantial numerical and analytical evidence indicates that ETH is robust in short-range interacting systems 
with chaotic spectral properties \cite{Alessio2016}.

This standard picture is however notably challenged by long-range interacting systems. Interactions decaying algebraically with distance, $r^{-\alpha}$, are known to give rise to phenomena absent in 
short-range systems. Atypical relaxation processes and anomalous transport have been observed in systems with sufficiently slowly decaying interactions \cite{Defenu2023,Defenu2024}. Other phenomena supported by long-range interacting systems include time crystals \cite{Zhang2017,Choi2017,Collura2022}, statistical ensemble inequivalence \cite{Russomanno2021}, prethermalization and the emergence of long-lived stationary states \cite{Halimeh2017b,Neyenhuis2017,Defenu2021,Mattes2025}, dynamical quantum phase transitions \cite{Zhang2017b,Halimeh2017,Homrighausen2017,Zunkovic2018,Piccitto2019,Ranabhat2022,Corps2025arxiv}, condensed-matter analogues of quark confinement \cite{Hauke2013,Lerose2019,Liu2019,Tan2021} or dynamical hadron formation \cite{Verdel2020,Vovrosh2022}. These features suggest a nontrivial modification of 
thermalization mechanisms and raise fundamental questions about the thermalization mechanism in the presence of long-range interactions.

The breakdown of conventional thermodynamic behavior becomes particularly severe when the interaction power-law exponent is $\alpha < d$, where $d$ is the spatial dimension. 
In this regime, the interaction energy is superextensive and the thermodynamic limit is ill defined. 
A standard remedy is the introduction of a Kac normalization factor \cite{Kac1963} into the Hamiltonian, which restores extensitivity by rescaling the 
interaction strength with system size. 
While this procedure regularizes the equilibrium properties, the resulting dynamics remains highly nontrivial.

In particular, in the collective limit $\alpha = 0$, the system admits a classical description 
characterized by an effective Planck constant $\hbar_{\mathrm{eff}} \sim 1/N$, with $N$ being the number of particles. These collective models serve as convenient testbeds for very long-ranged behavior. They are characterized by collective dynamics, essentially controlled by classical mean-field degrees of freedom and with quantum fluctuations suppressed by system size \cite{Cejnar2021}. Despite their seemingly simple structure, they exhibit a vast range of effects, including the excited-state quantum phase transition (ESQPT) \cite{Caprio2008,Stransky2014}, huge decoherence \cite{Relano2008,Perez2009}, 
singularities in quench dynamics \cite{Perez2011,Santos2015,Lobez2016,Bernal2017,Kloc2018}, feedback control in dissipative systems \cite{Kopylov2015}, quantum work
statistics \cite{Wang2017}, and localization \cite{Santos2016}, symmetry-breaking equilibrium states and dynamical phase transitions \cite{Puebla2013,Puebla2014,Corps2022,Corps2022prl,Corps2024prr,Corps2023b,Corps2024constants,Corps2021prl}, universal dynamical scaling \cite{Puebla2020}, dynamical instabilities \cite{Bastidas2014}, irreversibility without energy dissipation \cite{Puebla2015}, and reversible quantum information spreading \cite{Hummel2019}. The quantum-classical nature of these systems makes them particularly suitable to analyze quantum and classical thermalization mechanisms and chaos \cite{Cejnar2021,Sinha2024}.

Within this semiclassical framework, one expects a well-defined hierarchy of dynamical timescales. 
At short times, the quantum evolution of an initially coherent state follows the corresponding \textit{classical}
trajectory \cite{Chirikov1988}. 
In the presence of classical chaos, this correspondence persists up to the Ehrenfest time 
$\tau_{\mathrm{E}} \sim \ln N$ \cite{Chirikov1981,Berman1978,Berry1979}. 
Beyond this timescale, the wavepacket spreads over the classically accessible phase space, and its subsequent 
evolution is expected to be governed by an effectively classical diffusion process. 
This diffusive regime is anticipated to persist up to a different time scale, the Heisenberg time $\tau_{\mathrm{H}} \sim N^{a}$, where 
$a > 0$ depends on the spectral properties of the system \cite{Chirikov1981}. 
As a consequence, one would expect the quantum state to explore the available phase space, leading to equilibration dynamics similar to those already found in collective models such 
as the Dicke or Lipkin-Meshkov-Glick models in the appropriate energy regimes (see, e.g., \cite{Cejnar2021,LewisSwan2021,Lobez2016,Puebla2013,Corps2022,Corps2022prl,LermaHernandez2019})

In this work, we demonstrate that this semiclassical expectation can fail in a collective model. 
We present an explicit example in which the quantum wavepacket does not fully delocalize over the classical 
phase space, even at long times, and retains a substantial degree of coherence. 
We argue that this behavior originates from symmetry-breaking effects and the presence of quasi-degenerate 
eigenlevels, which effectively constrain the dynamics. Our numerical results are done for the 4-site Bose-Hubbard model; we also briefly comment on the effect of increasing the number of sites in the model. Our main result is an effective breakdown of the quantum-classical correspondence up to a large number of particles: the classical and quantum equilibrium values in a certain spectral region do not coincide. More generally, we find three distinct dynamical phases: (i) a low-energy region characterized by quasi-integrability and symmetry-breaking equilibrium states; (ii) a spectral region at intermediate energies characterized by ergodicity but where eigenstate expectation values of observables and classical averages do not coincide; and (iii) a high-energy region where the dynamics is ergodic and quantum and classical values coincide already for low numbers of particles. 

This paper is organized as follows. In Sec. \ref{sec:model} we review the collective Bose-Hubbard model, including its symmetries, as well as the main physical quantities that we measure in the rest of the paper. The classical limit of the model and relevant quantum operators are also presented in this section. The corresponding phase space structures as well as a quantum analogy for them are presented in Sec. \ref{sec:phasespaceanalysis2} and Sec. \ref{sec:quantumanalogy}, respectively. The quantum-classical correspondence is further analyzed in terms of the onset of chaotic behavior in Sec. \ref{sec:chaos}. We consider nonequilibrium dynamics, thermalization, and the nature (symmetric or symmetry-breaking) of such equilibrium states in Sec. \ref{sec:equilibrationdynamics}. Our conclusions are gathered in Sec. \ref{sec:conclusions}.

\section{Model}\label{sec:model}
We are interested in a model where the quantum-classical correspondence can be analyzed in a clear way. For this reason, and others we highlight below, we have considered the $\mathcal{N}-$site Bose-Hubbard model with periodic boundary conditions, whose Hamiltonian is given by
\begin{equation}
  \label{eq:model}
  \hat{H} = \sum_{i=1}^{\mathcal{N}}  \left[ -J \left( \hat{a}^{\dagger}_{i+1} \hat{a}_i + \hat{a}_{i+1} \hat{a}^{\dagger}_i \right) + \frac{U}{N} \hat{a}^{\dagger}_i \hat{a}^{\dagger}_i \hat{a}_i \hat{a}_i \right].
\end{equation}
The previous equation describes an entire family of models with $\mathcal{N}$ sites with $N$ bosonic particles, and $\hat{a}^{\dagger}_i$ and $\hat{a}_i$ are the usual creation and annihilation bosonic operators acting on the $i$th site as $\hat{a}_{i}\ket{n_{i}}=\sqrt{n_{i}}\ket{n_{i}-1}$ and $\hat{a}^{\dagger}_{i}\ket{n_{i}}=\sqrt{n_{i}+1}\ket{n_{i}+1}$. The hopping between the various sites is set to be constant, $J$, and $U$ is the interaction potential. 

This $\mathcal{N}$-site Bose-Hubbard model with $N$ bosons is a collective system \cite{Cejnar2021} and, as such, for $\mathcal{N}$ fixed, the infinite-size limit, $N\to\infty$, is mathematically equivalent to the classical limit, $\hbar_{\textrm{eff}}\propto 1/N\to 0$ (see below for a detailed computation of this limit in terms of collective positions and momenta). Thus, the classical limit is approached as a function of $N$. This property is independent of the number of sites as well as the topology and geometry of the lattice. We note that we do not consider here the version of the Bose-Hubbard model where the thermodynamic limit is approached by keeping constant the filling factor $N/\mathcal{N}$, in which case the thermodynamic and classical limits do not coincide.

 The Hamiltonian \eqref{eq:model} conserves the total number of particles, $[\hat{H},\hat{N}]=0$ with $\hat{N}=\sum_{i}\hat{a}_{i}^{\dagger}\hat{a}_{i}$. We consider periodic boundary conditions (PBCs). Some physical features of this model have been studied in different settings, with the number of sites typically set to $\mathcal{N}=3$. Some examples include persistent currents and vortex formation \cite{Tsubota2000,Scherer2007}, fragmented condensation \cite{Gallemi2015}, as well as a transition from regular to chaotic behavior \cite{Garcia-March2018, delaCruz2020,Wozniak2022,Nakerst2023} and its relation to symmetry-breaking and thermalizing eigenstates \cite{Gomez2025}. The Bose-Hubbard model is also a well-known minimal model for superfluid circuits \cite{Arwas2014,Arwas2017}. Although most of our numerical calculations are done for the model with $\mathcal{N}=4$ sites, the following discussion will be general for completeness. Model parameters will be set to $J=1$ and $U=-10$. 

The symmetries of the model are mathematically described by the dihedral group $D_{\mathcal{N}}$, corresponding to the transformations that leave the regular polygon of $\mathcal{N}$ sides invariant. All elements of this discrete group can be obtained from the following two non-commuting operators. First, a $2\pi/\mathcal{N}$ rotation, $\hat{R} \ket{n_1,\ldots,n_{\mathcal{N}-1},n_{\mathcal{N}}} = \ket{n_{\mathcal{N}},n_{1},\ldots,n_{\mathcal{N}-1}}$, where $n_i$ indicates the number of particles on site $i$. Since $\hat{R}^{\mathcal{N}}=1$, the eigenvalues of $\hat{R}$ are the $\mathcal{N}$th roots of unity, $r=e^{2k\pi i/\mathcal{N}}$, with $k=0,1,\ldots,\mathcal{N}-1$. Second, the model is invariant under a $\mathbb{Z}_{2}$ reflection transformation, $\hat{S} \ket{n_1,n_{2},\ldots,n_{\mathcal{N}-1},n_\mathcal{N}} = \ket{n_\mathcal{N},n_{\mathcal{N}-1},\ldots, n_{2},n_1}$ with two eigenvalues $s\in\left\lbrace 1, -1 \right\rbrace$. In general $[\hat{R},\hat{S}]\neq 0$, though exceptions apply to certain symmetry sectors. Thus Eq. \eqref{eq:model} can be diagonalized in the eigenbasis of $\hat{R}$, allowing us to classify the energy eigenstates according to the $\hat{R}$ quantum numbers as $\hat{H}\ket{E_{n,r}}=E_{n,r}\ket{E_{n,r}}$ with $r=0,\pi/2,\pi,3\pi/2$ in the case of $\mathcal{N}=4$. 

As a probe for the existence of symmetry-breaking phases in the system we use the generalized imbalance operator, 
\begin{equation}\label{eq:imbalanceop}
\hat{\mathcal{I}}=\sum_{k=1}^{\mathcal{N}}\hat{n}_{k}e^{2\pi i(k-1)/\mathcal{N}},\,\,k=1,\ldots,\mathcal{N},
\end{equation}
where $\hat{n}_{k}=\hat{a}_{k}^{\dagger}\hat{a}_{k}$. To facilitate comparisons with the classical limit (see below), we define the corresponding intensive operator as $\hat{I}=\hat{\mathcal{I}}/N$. Note that $\langle \hat{I}\rangle=e^{2\pi i(k-1)/\mathcal{N}}$ if all particles are located on the $k$th site. Because $\langle \hat{I}\rangle=0$ in the eigenstates of $\hat{R}$, it acts as a good order parameter in the case of ordered phases where the $2\pi/\mathcal{N}$ rotation symmetry is broken. 

For our purposes, a relevant aspect of this system is that it is a collective model with a well-defined classical counterpart. The classical limit can be computed by substitution of the bosonic quadratures $\hat{a}_{j} = \sqrt{N/2} \left( \hat{q}_{j} + i \hat{p}_{j} \right)$, and then taking the limit $N\to\infty$. This is equivalent to setting up an effective Planck constant $\hbar_{\textrm{eff}}\propto 1/N$, so the infinite size limit $N\to\infty$ coincides with the classical limit $\hbar_{\textrm{eff}}\to 0$. As a result, we obtain a classical, size-independent, energy functional $H(\mathbf{q},\mathbf{p})$ with $\mathcal{N}$ positions $q_i$, and momenta, $p_i$, defined by the limit $\lim_{N\to\infty}\hat{H}/N \mapsto H(\mathbf{q},\mathbf{p})$ and given by 
\begin{equation}\label{eq:classicalH}
    H(\mathbf{q},\mathbf{p})=\sum_{i=1}^{\mathcal{N}}\left[-J(q_{i+1}q_{i}+p_{i+1}p_{i})+\frac{U}{4}(q_{i}^{2}+p_{i}^{2})^{2}\right].
\end{equation}
Here, $(\mathbf{q},\mathbf{p})$ are canonical classical variables verifying $\{q_{i},p_{i}\}=1$ and $\{q_{i},p_{j}\}=\delta_{ij}$. However, due to particle number conservation, $\sum_{i=1}^{\mathcal{N}} q_i^2 + p_i^2=2$, the number of classical degrees of freedom is reduced to $\mathcal{N}-1$, with a classical phase space, $\Omega=\{(\mathbf{q},\mathbf{p})\in\mathbb{R}^{2\mathcal{N}}\,/\,\sum_{k=1}^{\mathcal{N}}q_{k}^{2}+p_{k}^{2}=2\}\subset \mathbb{R}^{2\mathcal{N}-1}$. In order to enable an easy comparison between classical ($N\to\infty$) and finite-$N$ quantum results, in this paper we use the intensive energy scale defined by $\epsilon\equiv E/N$, where $E$ denotes the actual (extensive) energy eigenvalues of Eq. \eqref{eq:model}. The infinite size limit of the model can be analyzed through the set of $2\mathcal{N}$ classical Hamilton equations 
\begin{equation}\label{eq:hameqs}
    \frac{\textrm{d}q_{i}}{\textrm{d}t}=\frac{\partial H}{\partial p_{i}},\,\,\,\frac{\textrm{d}p_{i}}{\textrm{d}t}=-\frac{\partial H}{\partial q_{i}},\,\,\,i=1,\ldots,\mathcal{N}.
\end{equation}
Similarly, the particle number operators on site $i$, $\hat{n}_{i}=\hat{a}_{i}^{\dagger}\hat{a}_{i}$, admit the classical representation 
\begin{equation}\label{eq:nclassical}
\lim_{N\to\infty}\frac{\hat{n}_{k}}{N}\mapsto n_{k}(\mathbf{q},\mathbf{p})=\frac{q_{k}^{2}+p_{k}^{2}}{2},\,\,k=1,2,\ldots,\mathcal{N}. 
\end{equation}
The generalized imbalance has the classical counterpart 
\begin{equation}\label{eq:imbalanceclassical}
    \lim_{N\to\infty} \frac{\hat{I}}{N}\mapsto I(\mathbf{q},\mathbf{p})=\frac{1}{2}\sum_{k=1}^{\mathcal{N}}(q_{k}^{2}+p_{k}^{2})e^{2\pi i(k-1) /\mathcal{N}}.
\end{equation}
To analyze the effects that the static structure of the model has on the thermalization procedure and the quantum-classical correspondence, we have calculated classical and quantum measures of the chaoticity of the system as well as the evolution of the occupation number operators, which can be used to determine whether an initial state evolves towards a symmetric or symmetry-breaking equilibrium state. We also observe a very slow scaling up to the thermodynamic limit for the numerically accessible number of particles that we can study. In what follows, we first study static properties of the model, followed by a dynamical analysis of chaos, and finally we study the equilibration process in the last subsection.

\section{Quantum and classical static features}\label{sec:phasespaceanalysis}

\subsection{Phase space analysis}\label{sec:phasespaceanalysis2}
In the case of collective models, where classicality is approached as the infinite-size limit parameter, $N$, is increased, some properties of the quantum behavior can be understood already at the classical level. Since our focus is on the equilibration value of physically relevant observables, and thus the long-time dynamics of such quantities, the behavior of classical trajectories in the classical limit offers valuable information. This strategy has been proved useful in other collective models, such as in the case of the Dicke model \cite{Puebla2013}, the Lipkin-Meshkov-Glick model \cite{Corps2022,Corps2023,Corps2023b}, the Rabi model \cite{Corps2021} and others \cite{Cejnar2021}, often with consequences such as dynamical quantum phase transitions \cite{Marino2022,Heyl2017}. The underlying reason is that the geometry and topology of the phase space can in some cases provide solid intuition about the behavior of classical trajectories, which for collective models is usually mirrored in the quantum domain.

One of the most impactful effects affecting the formation of phase space structures is the appearance of an ESQPT \cite{Cejnar2021}. An ESQPT is loosely defined as an extension of the QPT concept to the high-lying excited states of a quantum system, where the non-analytic behavior occurs at a given critical energy when all control parameters have been previously fixed. They arise from stationary points of the classical Hamiltonian, $\nabla H(\mathbf{q},\mathbf{p})=0$, and the type of non-analyticity depends on the index of the corresponding Hessian matrix \cite{Stransky2016}. For systems with $f$ classical effective degrees of freedom and non-singular Hessian matrix, the singular behavior induced by the ESQPT can be observed in the $(f-1)$th derivative of the level density, $\partial^{f-1}\rho(E)/\partial E^{f-1}$. This means that for a single degree of freedom a singularity can be observed in the level density itself. However, for systems with an exponentially large number of degrees of freedom, the spectral non-analyticity is smoothed out, as it is only revealed by increasingly high-order derivatives of the level density. Yet, the effects of the ESQPT, though not directly visible in the level density, can persist in the form of certain \textit{dynamical} effects in these systems (see, e.g., \cite{Corps2024prr,PerezFernandez2017} for explicit examples). Because we are interested in the BH model with $\mathcal{N}>2$ (and most of our numerical results are for $\mathcal{N}=4$), which features $\mathcal{N}-1$ effective degrees of freedom, our focus will be on the dynamics rather than on the particular shape of the level density. 

The ESQPT non-analyticity can trigger qualitative changes in the phase space geometry, affecting the system's dynamical behavior and thus the long-time average of physical quantities, which account for equilibrium values. The calculation of critical points $(\mathbf{q}_{c},\mathbf{p}_{c})$ leading to ESQPT critical energies $\epsilon_{c}=H(\mathbf{q}_{c},\mathbf{p}_{c})$ can become cumbersome if tackled by brute force methods, particularly for higher numbers of degrees of freedom. For the case of $\mathcal{N}=4$ this calculation can be successfully performed by making use of the Lagrange multiplier method developed in \cite{Novotny2025}. ESQPT critical points thus correspond to critical points of the Lagrange function 
\begin{equation}\label{eq:lagrange}
L(\mathbf{q},\mathbf{p},\lambda)=H(\mathbf{q},\mathbf{p})+\lambda\left[\sum_{k=1}^{\mathcal{N}}(q_{k}^{2}+p_{k}^{2})-2\right],
\end{equation}
for $\mathcal{N}=4$, $J=1$ and $U=-10$ in Eq. \eqref{eq:classicalH}. The constraint on the right-hand side of Eq. \eqref{eq:lagrange} comes from total particle number conservation, $\hat{N}=\sum_{k=1}^{\mathcal{N}}\hat{n}_{k}\mapsto \sum_{k=1}^{\mathcal{N}}q_{k}^{2}+p_{k}^{2}=2$. Critical points are solutions of $\nabla L(\mathbf{q},\mathbf{p})=0$. For our case, this procedure yields nine different critical energies: $\epsilon_{c}=-6.1,-5.4,-5,-4.824,-4.5,-4.1,-3.283,-2.5$. However, not all of these imply substantial changes in the system's dynamics. The two other stationary points are $-10.1$ and $-0.5$, but these correspond to the minimum and maximum energy of the system, respectively. 

\begin{figure}[h!]
\hspace*{-0.6cm}\includegraphics[width=0.46\textwidth]{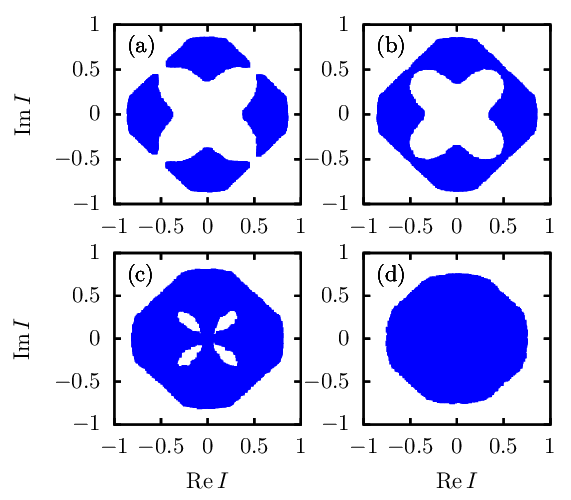}
    \caption{Real and imaginary parts of the classical generalized imbalance function, $I(\mathbf{q},\mathbf{p})$, in Eq. \eqref{eq:imbalanceclassical} for initial conditions $(\mathbf{q},\mathbf{p})\in\Omega$ at different energies, $\epsilon=H(\mathbf{q},\mathbf{p})$. (a) $-6.17\leq \epsilon\leq -6.13$. (b) $-6.07\leq \epsilon\leq -6.03$. (c) $-5.37\leq \epsilon\leq -5.33$. (d) $-4.57\leq \epsilon\leq -4.53$. All results are for $\mathcal{N}=4$, $J=1$ and $U=-10$.  }
    \label{fig:imbalances}
\end{figure}

In order to gain physical insight into this question, we seek to obtain some representation of the phase space $\Omega$. However, in the case of $\mathcal{N}=4$ sites, $\Omega$ is $7$-dimensional, which makes the naïve representation of constant energy surfaces futile (see \cite{Cejnar2021}, and many others, for examples). However, the generalized imbalance Eq. \eqref{eq:imbalanceop}, which is an operator connected to the symmetry-breaking in the rotation operator $\hat{R}$, can be used as an illustrative alternative. As we will see later, the picture this operator provides is useful to understand the equilibration effects we analyze. 

In Fig. \ref{fig:imbalances} we have represented the real and imaginary parts of the classical generalized imbalance in Eq. \eqref{eq:imbalanceclassical}. These are static results obtained by randomly sampling the classical constant energy surfaces, $\Omega_{E}=\{(\mathbf{q},\mathbf{p})\in\Omega\,/\,H(\mathbf{q},\mathbf{p})=E\}$, and then computing the generalized imbalance, $I(\mathbf{q},\mathbf{p})$, from the coordinates values $(\mathbf{q},\mathbf{p})$. To get a clearer picture for each panel we represent the imbalance in a narrow window of energies. We find four qualitatively different regions: 

(a) Fig. \ref{fig:imbalances}(a) shows the case for $-6.17\leq \epsilon\leq -6.13$, which is below all ESQPT critical energies mentioned above. The main feature of this case, which is general for any $\epsilon\leq \epsilon_{c1}=-6.1$, is the appearance of four disconnected wells. This suggests that classical trajectories $I(t)=I(\mathbf{q}(t),\mathbf{p}(t))$ that are initialized within one of the wells will remain trapped within that well for any time and, as a consequence, the classical equilibrium value,
\begin{equation}\label{eq:longtimeclassical}
\overline{I}\equiv \lim_{\tau\to\infty}\frac{1}{\tau}\int_{0}^{\tau}\textrm{d}t\,I(t),
\end{equation}
will be non-zero, $\overline{I}\neq 0$. This leads to a symmetry-breaking (SB) scenario where the proportion of particles on every site of the BH is not the same, $n_{i}\neq 1/\mathcal{N}$. This scenario undergoes a qualitative change upon crossing the first ESQPT at $\epsilon_{c1}=-6.1$.

(b) Indeed, in Fig. \ref{fig:imbalances}(b), where we display the case for $-6.07\leq \epsilon\leq -6.03$, the four wells become connected and there is a single hole that becomes inaccessible. In this case, however, a single trajectory may have access to the connected regions of the phase space, regardless of where it is initialized, and in the infinite-time limit it give rise to a symmetric state where $\overline{I}=0$ and so all sites become equally populated, $n_{i}=1/\mathcal{N}$. We would like to draw attention to the fact that the four wells are connected by quite narrow \textit{bridges}, and they become narrower as the ESQPT critical energy $E_{c1}$ is approached from above. This fact has important consequences in the quantum-classical correspondence of the model, as we will show later on. 

(c) In Fig. \ref{fig:imbalances}(c) we focus on the case $-5.37\leq \epsilon\leq -5.33$, which is above the second ESQPT at $\epsilon_{c2}=-5.4$. The four wells are again connected but we find four holes instead of only one as in the previous case. Classical trajectories will still be symmetric in general, as this only depends on the overall (dis)connection of the wells.

(d) Finally, in Fig. \ref{fig:imbalances}(d) we represent the case $-4.57\leq \epsilon\leq -4.53$. There are no longer four wells, but just one, and also all previous holes have disappeared as well, leading to a simpler symmetric scenario.

\begin{figure}[h!]
\hspace*{-0.6cm}\includegraphics[width=0.51\textwidth]{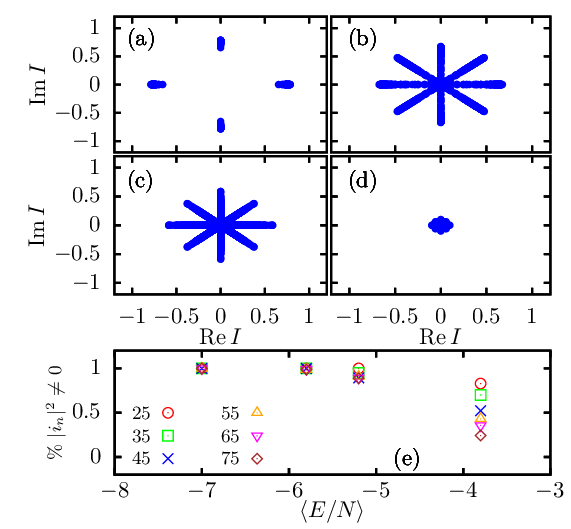}
    \caption{(a)-(d) Real and imaginary parts of the generalized imbalance operator, Eq. \eqref{eq:imbalanceop}, diagonalized in the respective energy eigenspaces, for $J=1$, $U=-10$ and $N=75$. The eigenvalues $\epsilon=E/N$ are chosen in the following energy ranges: (a) $-7.1\leq \epsilon\leq -6.9$. (b) $-5.9\leq \epsilon\leq -5.7$. (c) $-5.3\leq \epsilon\leq -5.1$. (d) $-3.9\leq \epsilon\leq -3.7$. (e) Number of eigenvalues with non-zero absolute value for different $N$. The threshold $|i_{n}|^{2}>10^{-3}$ is taken for the numerical computation. }
    \label{fig:imbalances_quantum}
\end{figure}

\subsection{Quantum analogy}\label{sec:quantumanalogy}
In order to analyze how these classical results compare with the behavior of the quantum model, in Fig. \ref{fig:imbalances_quantum} we have represented the real and imaginary parts of the generalized imbalance operator, Eq. \eqref{eq:imbalanceop}, for different energy ranges. To understand this result, we rely on the generalization of the ETH presented in \cite{Gomez2025} for the case of systems with symmetry-breaking eigenstates. Let $\hat{I}^{(n)}_{\alpha\beta}$ denote the reduced, $4\times 4$ imbalance operator written in the basis $\mathcal{B}_{n}=\left \{\ket{E_{n,0}},\ket{E_{n,\pi/2}},\ket{E_{n,\pi}},\ket{E_{n,3\pi/2}}\right \}$, spanned by the eigenstates of $\hat{H}$, $\{\ket{E_{n,r}}\}$, with different rotational quantum numbers, $r$, but the same $n=1,2,\ldots$ index, labelling the eigenspace number. The eigenvalues of $\hat{I}^{(n)}_{\alpha\beta}$ will be denoted $i_{n,r}$.  Similarly, the eigenvectors of $\hat{I}^{(n)}_{\alpha\beta}$ are denoted $\left \{\ket{\widetilde{E}_{n,r}}\right \}_{n,r}$ with $r=0,\pi/2,3\pi/2,\pi$, and let $\widetilde{\mathcal{B}}_{n}\equiv \textrm{span}\{\ket{\widetilde{E}_{n,r}}\}_{r}$ be the eigenspace spanned by these eigenstates. In the case of complete energy degeneracy between eigenvalues of different rotational quantum numbers and fixed $n$, $E_{n,r}=E_{n,r'}$ for all $r,r'$, the eigenvectors of $\hat{I}^{(n)}_{\alpha\beta}$ are trivially also eigenvectors of $\hat{H}$. However, if degeneracies are not exact, the eigenvectors of $\hat{I}^{(n)}_{\alpha\beta}$ are instead a linear combination of the eigenvectors of $\hat{H}$. 

In what follows, let us first consider the first case of exact degeneracy. This assumption is realistic in the case of systems with discrete $\mathbb{Z}_{n}$ symmetry-broken phases, where the gaps $\Delta_{n,rr'}\equiv |E_{n,r}-E_{n,r'}|$ close exponentially fast with $N$. In the symmetric phase the gap also decreases with $N$ as $1/N^{2}$. Let us consider a generic initial state, $\ket{\Psi(0)}$, written in the $\widetilde{\mathcal{B}}_{n}$ eigenbasis, 
\begin{equation}
\ket{\Psi(0)}=\sum_{n,r}c_{n,r}\ket{\widetilde{E}_{n,r}},
\end{equation}
with $c_{n,r}=\bra{\widetilde{E}_{n,r}}\ket{\Psi(0)}$. Allowing $\ket{\Psi(0)}$ to evolve under $\hat{H}$ results in the instantaneous wavefunction
\begin{equation}\label{eq:psit}
\ket{\Psi(t)}=\sum_{n,r}c_{n,r}e^{-i E_{n,r}t/\hbar}\ket{\widetilde{E}_{n,r}}.
\end{equation}
The time expectation value of the generalized imbalance under Eq. \eqref{eq:psit} is
\begin{equation}
\begin{split}
&\langle \hat{I}(t)\rangle=\sum_{n,r}|c_{n,r}|^{2}\bra{\widetilde{E}_{n,r}}\hat{I}\ket{\widetilde{E}_{n,r}}+\\ 
& \sum_{n\neq n'\oplus r\neq r'}c_{n',r'}^{*}c_{n,r}e^{-i(E_{n,r}-E_{n',r'})t/\hbar}\bra{\widetilde{E}_{n',r'}}\hat{I}\ket{\widetilde{E}_{n,r}}+\textrm{H.c.},
\end{split}
\end{equation}
where the notation $\oplus$ means that the conditions $n=n'$ and $r=r'$ cannot take place simultaneously (exclusive `or' operation), as this would produce the diagonal term in the first line above. The long-time average of $\langle \hat{I}\rangle$, representing the equilibration dynamics,
\begin{equation}\label{eq:lta}
\overline{\langle\hat{I}(t)\rangle}=\lim_{\tau\to\infty}\frac{1}{\tau}\int_{0}^{\tau}\textrm{d}t\,\langle \hat{I}(t)\rangle,
\end{equation}
reaches the asymptotic value
\begin{equation}\label{eq:lta2}
\begin{split}
\overline{\langle \hat{I}(t)\rangle}&=\sum_{n,r}|c_{n,r}|^{2}\bra{\widetilde{E}_{n,r}}\hat{I}\ket{\widetilde{E}_{n,r}}+\\ 
& \sum_{n,r\neq r'}c_{n,r'}^{*}c_{n,r}\bra{\widetilde{E}_{n,r'}}\hat{I}\ket{\widetilde{E}_{n,r}}+\textrm{H.c.}
\end{split}
\end{equation}
Two reasons sustain this last equation. First we have the exponentials with Bohr frequencies involving $E_{n,r}$ and $E_{n,r'}$ with $r\neq r'$. Due to the assumption of degeneracies, these frequencies vanish, $E_{n,r}=E_{n,r'}$, which is responsible for the second constant term in Eq. \eqref{eq:lta2}. Second we have the frequencies with $E_{n,r}$ and $E_{n',r}$, which are not degenerate as these eigenvalues correspond to different symmetry sectors. These, however, do not survive the long-time average and therefore they do not contribute to Eq. \eqref{eq:lta2}. The eigenstate expectation values on the top line are precisely the eigenvalues $i_{n,r}=\bra{\widetilde{E}_{n,r}}\hat{I}\ket{\widetilde{E}_{n,r}}$, while the expectation values on the bottom line vanish because $r\neq r'$. Therefore, Eq. \eqref{eq:lta2} can be further simplified as
\begin{equation}\label{eq:lta3}
\overline{\langle \hat{I}(t)\rangle}=\sum_{n,r}|c_{n,r}|^{2}i_{n,r}.
\end{equation}
This is simply a diagonal ensemble \cite{Alessio2016} but written in the  $\widetilde{\mathcal{B}}_{n}$ eigenbasis. It tells us that for $\hat{I}$ the equilibration values merely depend on the interplay between the population coefficients $c_{n,r}$ and the imbalance eigenvalues $i_{n,r}$. If there are some $i_{n,r}\neq 0$, then symmetry-breaking is possible if there is a non-zero overlap with these states.

In Fig. \ref{fig:imbalances_quantum}(a) we focus on the eigenvalues coming from matrices projected onto the low-energy regime. The modulus of such eigenvalues is always different from zero, which is indicative of the presence of symmetry-breaking states in this region. In the intermediate above the first ESQPT shown in Figs. \ref{fig:imbalances_quantum}(b)-(c) we observe a significant number of eigenvalues that are still non-vanishing. This indicates the possibility of preparing long-lived symmetry-breaking initial states also in this region, even though classically it corresponds to a topologically connected phase. Finally, for even higher energies, away from the first ESQPT, Fig. \ref{fig:imbalances_quantum}(d) shows that most eigenvalues are close to zero, which is compatible with a dynamical phase dominated by symmetric states. Since these results are for $N=75$ particles, in Fig. \ref{fig:imbalances_quantum}(e) we present a finite-size analysis of the proportion of eigenvalues that are non-vanishing. For high energies this proportion decreases quite fast with $N$, while it seems to remain constant for very low energies close to the ground state of the system. Yet, for intermediate energies, in particular close but above the first ESQPT, the number of non-vanishing eigenvalues decreases only very slightly with $N$. This suggests that as the energy gap decreases as $1/N^{2}$ (on average, the number of Hamiltonian eigenlevels increases as $N^{3}$, and the energy is extensive and increasing as $N$), the proportion of eigenvalues of the generalized imbalance with large modulus barely decreases in this intermediate region, which seems compatible with a high relevance of symmetry-breaking states even for relatively high values of $N$. 

\begin{figure}[h!]
\hspace*{-0.45cm}\includegraphics[width=0.44\textwidth]{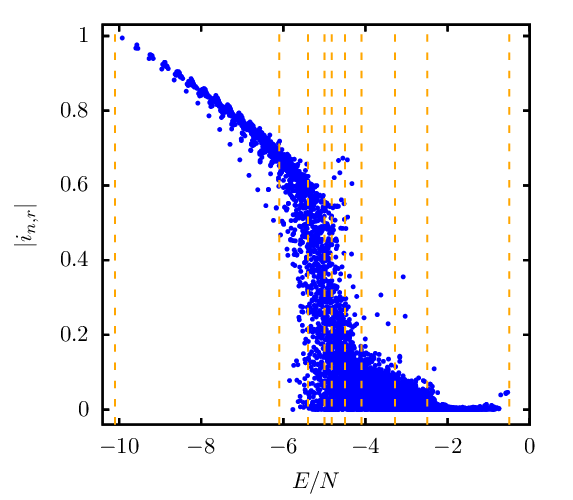}
    \caption{Eigenvalues of the generalized imbalance operator $\hat{I}$, defined in Eq. \eqref{eq:imbalanceop}, in the $4\times 4$ reduced eigenspaces $\{E_{n,r}\}$ with $r=0,\pi/2,\pi,3\pi/2$, as a function of energy, $i_{n}$. Dashed orange lines represent the various ESQPT critical energies, together with the ground-state (first line) and the maximum energies (last line). Model parameters are $J=1$, $U=-10$, and $N=55$. }
    \label{fig:aut_imbalance}
\end{figure}

In order to delve deeper into this question, in Fig. \ref{fig:aut_imbalance} we have represented the modulus of the eigenvalues $i_{n,r}$ of $\hat{I}^{(n)}$ as a function of energy. These have been obtained by writing the generalized imbalance $\hat{I}$ in each of the reduced $4\times 4$ energy subspaces, as in Fig. \ref{fig:imbalances_quantum}. We observe that for low energies below all ESQPTs, $\epsilon<-6.1$, all values are $i_{n,r}\neq 0$, indicating a symmetry-breaking scenario where $\langle \hat{I}\rangle \neq 0$. At very high energies, these eigenvalues are all $i_{n,r}=0$, corresponding to a symmetric situation with $\langle \hat{I}\rangle=0$. And there is also an intermediate region where most $i_{n,r}$ cluster around zero but also with many $i_{n,r}\neq 0$. This has important implications for the equilibration process, as we discuss below and explicitly exemplify in Sec. \ref{sec:equilibrationdynamics}.

Based on Eq. \eqref{eq:lta3}, we find three different situations:

(i) In the symmetry-breaking region at low energies, we have $i_{n,r}\neq 0$ for all $n$, and therefore all initial conditions will give rise to $\overline{\langle \hat{I}\rangle}\neq 0$, with $\overline{\langle \hat{n}_{k}\rangle}\neq 1/\mathcal{N}$ for all $k=1,2,\ldots,\mathcal{N}$.  This is because for symmetric equilibrium states to exist in this phase, with $\overline{\langle \hat{I}\rangle}=0$, macroscopic superpositions would be necessary (for example, the first site maximally populated and, at the same time, maximal population also at the second, third, and fourth sites).

(ii) In the symmetric region at high energies, all eigenvalues are $i_{n,r}\approx 0$, and therefore $\overline{\langle \hat{I}\rangle}=0$ irrespective of the distribution of $\{c_{n,r}\}_{n,r}$. This leads to $\overline{\langle \hat{n}_{k}\rangle}=1/\mathcal{N}$ for all $k=1,2,\ldots,\mathcal{N}$. 

(iii) In the intermediate region above the first ESQPT there is a coexistence of $i_{n,r}=0$ and $i_{n,r}\neq 0$, so the resulting long-time average depends on the set of $i_{n,r}$ such that the population coefficients $c_{n,r}$ are significantly different from zero. 

A technical note is warranted. This reasoning is a product of Eq. \eqref{eq:lta3} which is based on assuming exact degeneracy of eigenlevels within a single energy subspace $\mathcal{B}_{n}$, $\Delta_{n,rr'}=|E_{n,r}-E_{n,r'}|=0$ for all $r,r'$.  This degeneracy is not exact for $N<\infty$, but in the case of discrete $\mathbb{Z}_{\mathcal{N}}$ symmetry breaking it is realistic to assume that it is because the gaps close exponentially fast with $N$, in contrast with systems with continuous symmetry-breaking, where the gaps seem to close algebraically \cite{Khalouf2024}. Even in the symmetry-restored phase, the mean level spacing decays as $1/N^{2}$, which means that the results above remain approximately valid. In any case, given that $\Delta_{n,rr'}\neq 0$ in finite-size systems, this image only survives up to a certain timescale; for symmetry-breaking phases, it is of the order $\tau\sim 1/\Delta_{n,rr'}\sim e^{\alpha N}$ for some $\alpha>0$, which is extremely long and can exceed the age of the Universe given that at low energies $\Delta_{n,rr'}=\mathcal{O}(10^{-16})$ already for quite small values of $N$. For times longer than this scale, $\ket{\widetilde{E}_{n,r}}$ ceases to behave as a Hamiltonian eigenstate, and the derivation above is technically invalid. In the symmetric phase, this characteristic timescale is instead $\tau\propto N^{2}$.

The results of this section are based on the ergodic hypothesis, which implies that the ensuing equilibrium state depends merely on the properties of the classical phase space and that there are no constraints on the quantum population coefficients, meaning that there are no selection rules imposed by certain conserved quantities. In the following section we test if and where such an ergodic hypothesis actually holds for our system thorugh the analysis of quantum chaos.

\section{Classical and quantum chaos}\label{sec:chaos}

\begin{figure}[h!]
\hspace*{-0.9cm}\includegraphics[width=0.55\textwidth]{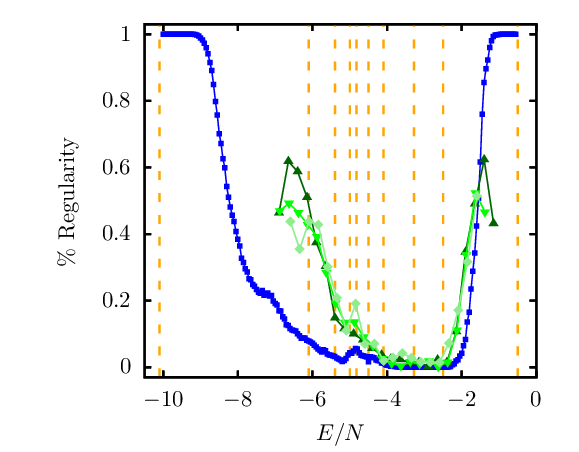}
    \caption{Quantitative analysis of chaos in the 4-site BH model. The blue line shows the classical fraction of regularity, $f_{\textrm{reg}}$, obtained from the Lyapunov exponent, Eq. \eqref{eq:lyapunov}, calculated for $10^{4}$ initial conditions at each given energy. The green lines shows the fraction of regularity as obtained from the Berry-Robnik distribution, Eq. \eqref{eq:psberryrobnik} for different system sizes: $N=75$ (light green), $101$ (green) and $125$ (dark green). Dashed orange lines represent the different ESQPT critical energies as well as the ground-state and maximum energy values. }
    \label{fig:chaos}
\end{figure}

The main mechanism underpinning the onset of thermal behavior is chaos. This is well known in the case of classical dynamics \cite{Gutzwiller1990}, where ergodic behavior ensures that phase space averages coincide with long-time averages of the form of Eq. \eqref{eq:longtimeclassical}. The main phenomenological idea of classical chaos is that of a dynamical regime characterized by an exponential divergence of two initial conditions that, initially, were infinitesimally close to each other. Regular, non-chaotic trajectories, can only feature a separation that grows algebraically at most. These ideas are most commonly quantified through the Lyapunov exponents \cite{Cencini2010}. Given an initial condition in phase space, $\mathbf{X}(0)\equiv (\mathbf{q}(0),\mathbf{p}(0))\in\Omega$, and a small perturbation, $\delta \mathbf{X}(0)$, chaos means that the perturbation on the trajectory is amplified as $\norm{\delta \mathbf{X}(t)}\sim \norm{\delta \mathbf{X}(0)}e^{\lambda_{L}t}$, where $\lambda_{L}\geq 0$ is the (largest) Lyapunov exponent, 
\begin{equation}\label{eq:lyapunov}
\lambda_{L}=\lim_{t\to\infty}\lim_{\norm{\delta \mathbf{X}(0)}\to 0}\frac{1}{t}\frac{\norm{\delta\mathbf{X}(t)}}{\norm{\delta\mathbf{X}(0)}}.
\end{equation}
This limit is well defined and it equals a finite number for almost any $\mathbf{X}(0)\in\Omega$ and $\delta\mathbf{X}(0)$ \cite{Osledec1986}. If $\lambda_{L}>0$, the trajectory is chaotic. In our numerical calculations we have used the tangent dynamics method to compute the Lyapunov exponent \cite{Parker1989}. We have selected $10^4$ initial conditions $(\mathbf{q},\mathbf{p})\in\Omega$ uniformly on a hypersphere with radius $\sqrt{2}$ (defined by the constraint $\sum_{k}q_{k}^{2}+p_{k}^{2}=2$) and energy nor further than $10^{-4}$ from a given selected energy. We have repeated this procedure for initial conditions at different energies in order to sample the entirety of phase space, and computed the average Lyapunov exponent in each case. We have also computed the fraction of regular trajectories, $f_{\textrm{reg}}$, defined as the ratio of the number of trajectories with vanishing Lyapunov exponent to the total number of trajectories considered in the ensemble. Numerically, one has to arbitrarily set a threshold below which the Lyapunov exponent is considered to be zero, which in this case is $0.01$. In the limit of an infinite number of trajectories, $f_{\textrm{reg}}$ is a measure of the volume of the regular part of classical phase space compared to the total phase space volume, and thus it provides information about the proportion of trajectories traped in classical torii. Such a fraction of regularity $f_{\textrm{reg}}$ as a function of the energy value is represented in Fig. \ref{fig:chaos} with blue points.

On the other hand, the signatures of chaos in the quantum domain are commonly used to define the concept of quantum chaos, which is in turn intimately connected with quantum thermalization and the ETH \cite{Alessio2016,Srednicki1994,Srednicki1999,Santos2010}. As stated in the famous quantum chaos conjecture \cite{Bohigas1984}, the level statistics of quantum systems whose classical counterpart is chaotic follow the universal description of the random matrix theory \cite{Mehta2004}. To measure the degree of quantum chaos in our system, we use the traditional nearest-neighbor level spacing distribution $P(s)$, where $s_{n}=\epsilon_{n+1}-\epsilon_{n}$, where $\{\epsilon_{n}\}_{n}$ denotes the unfolded eigenvalues (see below). In the case of semiclassical systems exhibiting the coexistence of chaotic and regular phase space regions, Berry and Robnik \cite{Berry1984} derived an expression for $P(s)$ assuming that eigenlevel sequences from chaotic regions follow the Wigner-Dyson distribution for quantum chaotic systems \cite{Mehta2004,Bohigas1984}, $P(s)=\frac{\pi s}{2}e^{-\pi s^{2}/4}$, while sequences from regular regions follow the Poisson distribution \cite{Berry1997}, $P(s)=e^{-s}$. The result is the following expression for $P(s)$ the case of mixed semiclassical dynamics,
\begin{equation}\label{eq:psberryrobnik}
    P(s;\rho)=\rho^{2}e^{-\rho s}\textrm{erfc}(\frac{1}{2}\sqrt{\pi}\overline{\rho}s)+(2\rho\overline{\rho}+\frac{1}{2}\pi\overline{\rho}^{3}s)e^{-\rho s-\frac{1}{4}\pi\overline{\rho}^{2}s^{2}},
\end{equation}
where $\textrm{erfc}(x)=\frac{2}{\sqrt{\pi}}\int_{x}^{\infty}\textrm{d}t\,e^{-t^{2}}$, $\rho$ is the fraction of Poissonian (regular) sequences and $\overline{\rho}=1-\rho$ is the fraction of chaotic trajectories. This approach is convenient to analyze the quantum-classical correspondence because $\rho$ represents the proportion of classical regular orbits. Equation \eqref{eq:psberryrobnik} only depends on the regularity parameter $\rho$, and thus it may be fitted to the numerically computed $P(s)$, obtained from a quantum spectrum, to extract its value.

 We have computed the level spacing distribution for the quantum model Eq. \eqref{eq:model} for $N=125$ particles, $J=1$ and $U=-10$, calculating the eigenvalues for the $r=0,\pi$ symmetry sectors, and separating the corresponding spectra according to the reflection transformation, $s=1,-1$. We have then considered the averaged $P(s)$ distribution calculated from these four level spacing distributions. To eliminate the spurious influence of the distribution binning, we have obtained $\rho$ through the cumulative distribution of Eq. \eqref{eq:psberryrobnik}, $F(s)=\int_{0}^{s}\textrm{d}s'\,P(s';\rho)$.  The universal random matrix theory predictions only hold on the so-called unfolded scale, where system-specific contributions to be density of states have been removed and $\langle s\rangle=1$. The unfolded energy values are computed by fitting the actual eigenvalues $\{E_{n}\}$ to a polynomial in order to simulate the cumulative level density $\epsilon_{n}=\int_{-\infty}^{E_{n}} \textrm{d}x\,\varrho(x)$, a procedure that has been described in detail in the literature \cite{Gomez2002,Corps2021pre}. In our case, the fitting polynomial is of degree 5. We have chacked that other polynomial degrees do not induce relevant qualitative changes. The result for $\rho$ is represented in Fig. \ref{fig:chaos}. Because the spectral density is quite low at low energy values, leading to insufficient statistics for the analysis, we have only considered energy partitions from $\epsilon\gtrsim -7$, which covers the first ESQPT. Also, the spectral statistics near the ground-state can show non-universal behavior due to the presence of quasi-conserved quantities, which makes the interpretation of level statistics complicated in a such low energy range.

As seen in Fig. \ref{fig:chaos}, the system dynamics is dominated by regularity close to the ground-state and maximum energies. This most probably due to the appearance of emergent conserved quantities in this region, similar to those described in \cite{Relano2016,Bastarrachea2017,Corps2022jpa} in a different setting. Most chaotic dynamics take place in the bulk of the spectrum, where the fraction of regularity sharply diminishes both classically and quantum mechanically. The agreement of the classical $f_{\textrm{reg}}$ and the semiclassical value of $\rho$ from Eq. \eqref{eq:psberryrobnik} is good in the deep chaotic regime $-4\lesssim \epsilon\lesssim -2$, with significant discrepancies clearly visible for $\epsilon \lesssim -5$. According to the value of $f_{\textrm{reg}}\sim 0.1$, the system is classically strongly chaotic at energies slightly above the first ESQPT at $\epsilon_{c1}=-6.1$; however, Berry-Robnik's fraction of regularity is quite high at around $60\%$. 

Let us now consider the case $\epsilon\gtrsim -6.1$. The accessible region of the classical phase space is the $6$-dimensional hypersurface defined by the intersection of the hypersurfaces $H(\mathbf{q},\mathbf{p})=E$ and $\sum_{k}p_{k}^{2}+q_{k}^{2}=2$, and it is topologically connected. In a mixed classical phase space region, there are two types of trajectories: ergodic or regular. Ergodic trajectories cover all regions in a connected phase space, while regular trajectories are restricted to torii. For this reason, a chaotic trajectory is necessarily symmetric in a connected phase space, as it respects the global Hamiltonian symmetry. It follows that any symmetry-breaking trajectory is necessarily trapped in a torus. Thus, given that $f_{\textrm{reg}}\sim 0.1$ around $\epsilon=-6$, we can conclude that the number of symmetry-breaking trajectories in this region is bounded from above by this percentage. This bound comes from the fact that any symmetry-breaking trajectory is trapped on a torus, but the converse is not necessarily true. Yet, the analysis of the eigenvalues of the generalized imbalance shown in Fig. \ref{fig:imbalances_quantum} shows a percentage that is significantly higher than $0.1$ in the same energy region, and this coincides with the estimation of regularity obtained from the Berry-Robnik distribution, which is also visibly higher than $0.1$. This scenario does not seem to change a lot with increasing number of particles, at least for the values of $N$ that we can realistically reach. This seems to suggest a very low scaling up to the thermodynamic limit, indicating that these dynamical and static properties remain relatively unchanged for modest system sizes, perhaps leading to finite-size effects that survive up to high numbers of particles.

\section{Equilibration dynamics}\label{sec:equilibrationdynamics}
The dynamical effects observed in this sytem are one of the main results of our work. To reveal these effects, we make use of three measurements. 

(1) We consider a single classical trajectory defined by coordinates $(\mathbf{q}(0),\mathbf{p}(0))\in\Omega$ at a given energy value, $E=H(\mathbf{q},\mathbf{p})$, and compute its time evolution, $(\mathbf{q}(t),\mathbf{p}(t))$, according to the Hamilton equations \eqref{eq:hameqs}. The classical evolution of the particle number operators are computed as prescribed by Eq. \eqref{eq:nclassical}. 

(2) In the case of collective models, the quantum mechanical infinite-size $N\to\infty$ dynamics converges to the classical behavior. To compare the behavior of classical trajectories, characterized by unique values of the coordinates $(\mathbf{q},\mathbf{p})$, with a quantum counterpart, we resort to the coherent state
\begin{widetext}
\begin{equation}\label{eq:coherentstate}
\ket{\Phi}=\frac{1}{M}\sum_{\substack{n_{1},n_{2},n_{3},n_{4}\geq 0,\\ n_{1}+n_{2}+n_{3}+n_{4}=N}}\frac{(q_{1}+ip_{1})^{n_{1}}}{\sqrt{n_{1}!}}\frac{(q_{2}+ip_{2})^{n_{2}}}{\sqrt{n_{2}!}}\frac{(q_{3}+ip_{3})^{n_{3}}}{\sqrt{n_{3}!}}\frac{(q_{4}+ip_{4})^{n_{4}}}{\sqrt{n_{4}!}}\ket{n_{1},n_{2},n_{3},n_{4}},
\end{equation}
\end{widetext}
where $M$ is a normalization factor such that $\bra{\Phi}\ket{\Phi}=1$. For sufficiently large values of $N$, the energy width of Eq. \eqref{eq:coherentstate} in the eigenstates of the Hamiltonian Eq. \eqref{eq:model} shrinks, approaching a classical situation where the population coefficients follow a delta distribution around the mean energy value, $\delta(E-\langle E\rangle)$. Thus, we expect the quantum dynamics of the coherent state to mimic the classical results for sufficiently high $N$. This coherent state is taken as an initial state, $\ket{\Phi}\equiv \ket{\Phi(t=0)}$, and then left to evolve under $\hat{H}$ through the time-independent Schrödinger equation,
\begin{equation}\label{eq:phit}
\ket{\Phi(t)}=e^{-i\hat{H}t/\hbar}\ket{\Phi(0)}=\sum_{n,r}c_{n,r}e^{-i E_{n,r}t/\hbar}\ket{E_{n,r}},
\end{equation}
with $c_{n,r}$ such that $\sum_{n,r}|c_{n,r}|^{2}=1$ denoting the population coefficients $c_{n,r}=\bra{E_{n,r}}\ket{\Phi(0)}$. In our calculations we set $\hbar=1$. 

(3) In the finite-size case $N<\infty$, the coherent state Eq. \eqref{eq:coherentstate} strictly has a finite spectral width, and thus it can be argued that it should be not directly compared with a single classical trajectory. In order to take into account the non-classicality of the coherent state, we have also computed an average over 2000 classical trajectories with initial conditions $(\mathbf{q}(0),\mathbf{p}(0))$ taken from a Gaussian distribution centered at the same coordinate values as the quantum coherent state, with variance $\sigma=1/\sqrt{2N}$, and in such a way that the phase space constraint $\sum_{k}q_{k}^{2}+p_{k}^{2}=2$ is respected (i.e., the truncated Wigner approximation, see e.g. \cite{Polkovnikov2010}). This average over classical trajectories should closely mimic the behavior of the coherent state. 

\begin{figure}[h!]
\hspace*{-0.45cm}\includegraphics[width=0.49\textwidth]{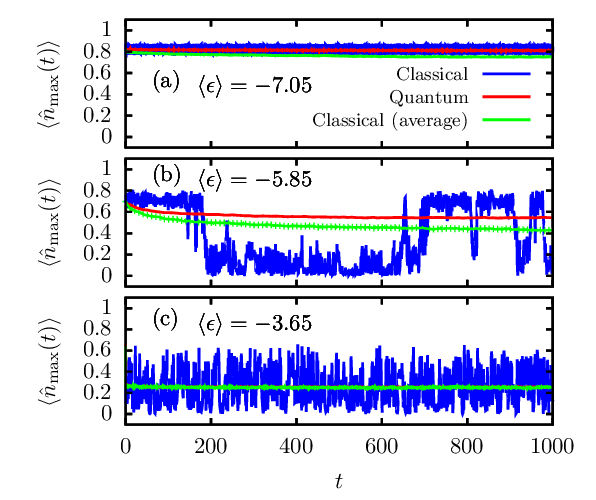}
    \caption{Time evolution of the occupation number for classical trajectories with initial conditions $(\mathbf{q},\mathbf{p})$ at fixed energies $\epsilon=H(\mathbf{q},\mathbf{p})$ defined by Eq. \eqref{eq:classicalH} (blue) and time evolution for the coherent state Eq. \eqref{eq:coherentstate} (red) with the same $(\mathbf{q},\mathbf{p})$. For each panel we represent the time evolution $\langle \hat{n}_{\max}(t)\rangle$ of the number particle operator of the well with the initially largest number of particles. The green lines represent the average of 2000 classical trajectories with initial coordinates $(\mathbf{q},\mathbf{p})$ taken from a Gaussian distribution centered at the same coordinate values as the quantum coherent states and with a width $\sigma=1/\sqrt{2N}$ with $N=55$. Three different cases are selected: (a) $\epsilon=-7.05<\epsilon_{c1}$ below the first ESQPT, with $(\mathbf{q},\mathbf{p})\approx ((0.086,-0.171,-0.9,0.151),(-0.38,0.392,0.911,-0.048))$, (b) $\epsilon=-5.85>\epsilon_{c1}$ slightly above the first ESQPT, with $(\mathbf{q},\mathbf{p})\approx ((0.211,0.393,-0.221,0.224),(-0.623,-0.022,0.045,-1.145))$, and (c) $\epsilon=-3.65$ significantly above it, with $(\mathbf{q},\mathbf{p})\approx ((0.306,-0.948,0.289,-0.01),(-0.58,0.652,-0.373,0.151))$. Model parameters are $J=1$, $U=-10$, $N=55$. }
    \label{fig:timeevo}
\end{figure}

\begin{figure}[h!]
\hspace*{-0.45cm}\includegraphics[width=0.48\textwidth]{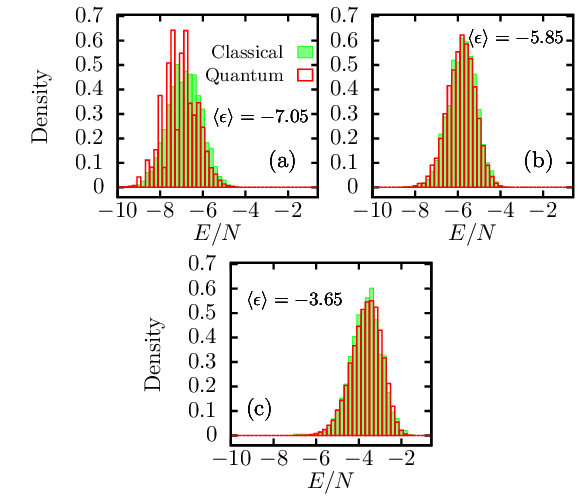}
    \caption{Local density of states for an average over 2000 classical trajectories with initial conditions chosen following a Gaussian distribution centered at (a) $\epsilon=-7.05<\epsilon_{c1}$ below the first ESQPT, with $(\mathbf{q},\mathbf{p})\approx ((0.086,-0.171,-0.9,0.151),(-0.38,0.392,0.911,-0.048))$, (b) $\epsilon=-5.85>\epsilon_{c1}$ slightly above the first ESQPT, with $(\mathbf{q},\mathbf{p})\approx ((0.211,0.393,-0.221,0.224),(-0.623,-0.022,0.045,-1.145))$, and (c) $\epsilon=-3.65$ significantly above it, with $(\mathbf{q},\mathbf{p})\approx ((0.306,-0.948,0.289,-0.01),(-0.58,0.652,-0.373,0.151))$ (filled green histograms) and the same result for the quantum coherent state Eq. \eqref{eq:coherentstate} defined by the same coordinates (red histograms). Model parameters are $J=1$, $U=-10$, and $N=55$ (for the quantum case). }
    \label{fig:distributions}
\end{figure}

In Fig. \ref{fig:timeevo} we have represented these three dynamical measurements in different spectral regions, $\langle \epsilon\rangle=-7.05,-5.85$ and $-3.65$. In all cases the initial condition is rotationally symmetry-broken because the number of particles in one of the four sites is significantly higher than in the remaining three sites.  The site with the highest number of particles changes with $\langle \epsilon\rangle$, so in each panel we have represented the number particle operator/function corresponding to the most populated site to compare our results more easily. This operator is denoted as $\hat{n}_{\max}$ in what follows. Figure \ref{fig:distributions}, where we show the corresponding local density of states (LDOS) for the quantum coherent state and the classical average of trajectories, complements these results. 

Let us first focus on the case $\langle \epsilon\rangle=-7.05$, which is below all ESQPTs and in a disconnected region of the classical phase space. Fig. \ref{fig:timeevo}(a) shows that the single classical trajectory, the coherent state, and the average over 2000 classical trajectories all reach a symmetry-breaking equilibration value as $n_{\max}\neq 1/\mathcal{N}=0.25$. This is to be expected from the qualitative shape of the classical phase space sketched in Fig. \ref{fig:imbalances}(a). Because the classical wells are disconnected, the initial state cannot explore the other regions and thus it will remain symmetry-broken. The overall behavior of the coherent state agrees quite well with the classical trajectory defined by the same coordinates. Here, the fluctuations for the coherent dynamics are suppressed due to the wavepacket dispersing in phase space, which leads to effective self-average. However, there are some visible discrepancies with the average over classical trajectories. This can be understood from the LDOS represented in Fig. \ref{fig:distributions}(a). We find that the classical and quantum LDOS are quite different, even though they lead to approximately the same average energy. While the classical distribution is smoother, there are some energy values for which the LDOS for the coherent state is particularly low. As mentioned in Sec. \ref{sec:chaos}, the region near the ground-state is possibly influenced by quasi-conserved quantities leading to quasi-integrability, and our set of trajectories has been built in a fully random way, without taking this possibility into account. 

Let us now consider the opposite case, $\langle\epsilon\rangle=-3.65$, depicted in Figs. \ref{fig:timeevo}(c) and \ref{fig:distributions}(c), and leave the intermediate case for last. This energy is far above the first ESQPT at $\epsilon_{c}=-6.1$, which is the one topologically relevant as it connects the classical phase space. As a consequence, all three dynamical measures lead to a symmetric equilibrium value with $n_{k}=0.25$. This is clearly observed in Fig. \ref{fig:timeevo}(c). Let us also notice that the evolution of the coherent state and that of the classical average coincide almost exactly. This is a direct consequence of the ergodicity of the system at this energy, also reflected in the Lyapunov exponents and the fraction of regularity in Fig. \ref{fig:chaos}. Similarly, the population distributions in Fig. \ref{fig:distributions}(c) show a very good agreement between the classical and quantum cases, which is in contrast to the results in Fig. \ref{fig:distributions}(a). 

\begin{figure}[h!]
\hspace*{-0.51cm}\includegraphics[width=0.51\textwidth]{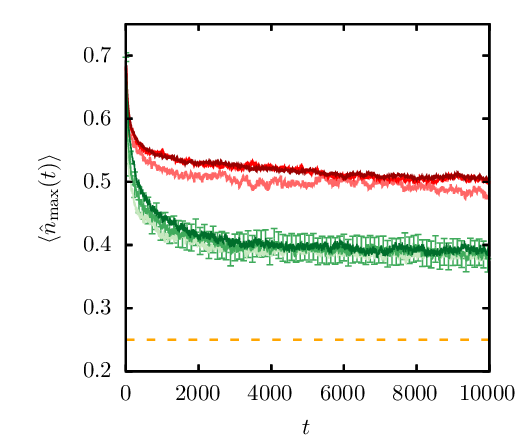}
    \caption{Scaling of the expectation value of the maximally populated site, $\hat{n}_{\max}$, for the quantum coherent state Eq. \eqref{eq:coherentstate} for the initial coordinates $(\mathbf{q},\mathbf{p})\approx ((0.211,0.393,-0.221,0.224),(-0.623,-0.022,0.045,-1.145))$. The state average energy is $\langle\epsilon\rangle=-5.85$. Model parameters are $J=1$, $U=-10$. The green line represents a classical average of trajectories defined with initial conditions $(\mathbf{q},\mathbf{p})$ taken from a Gaussian distribution centered at the same coordinate values as the quantum coherent states and with width $\sigma=1/\sqrt{2N}$. The results for the quantum coherent state (the classical average) are represented in red (green) with $N=35,55,75$ from ligther to darker color. Errorbars are represented in the classical average for $N=55$ and they correspond to three standard deviations from the mean. The symmetric value $\overline{n}_{k}=0.25$ is represented with orange dashed lines for clarity.}
    \label{fig:scaling_quantum}
\end{figure}

Finally, let us consider the case $\langle\epsilon\rangle=-5.65$ in Figs. \ref{fig:timeevo}(b) and \ref{fig:distributions}(b), where we observe the most interesting effects. First, we observe an intermittency behavior in the classical occupation number. Because $\langle\epsilon\rangle$ is above the first ESQPT, the phase space is connected, but only by very narrow classical bridges, as previously shown in Fig. \ref{fig:imbalances}(b). Thus a classical trajectory initialized in a given classical well is theoretically able to access the remaining three wells, and for sufficiently long times the long-time average should approach the symmetric equilibrium value $\overline{n}_{k}=0.25$. However, this expectation is not observed in Fig. \ref{fig:timeevo}(b), for the time values considered. Instead, we find an intermittent pattern in the time evolution of $n_{\max}(t)$, where it jumps quite suddenly between different values. This is a consequence of the tightness of the bridges connecting different regions of classical phase space, whereby the system gets effectively trapped in a single well as it struggles to find the paths to a different well. This effective trapping phenomenon means that the classically expected equilibrium value takes exceedingly long times to reach, in such a way that the system essentially seems out of equilibrium. Perhaps more striking is the observation that the quantum coherent state (red line) does not mimic, even remotely, the classical behavior in this case. It also does not approach the symmetric value $n_{k}=0.25$, reaching a relatively stable value that is quite far from it. The average over classical trajectories (green line) does not solve this discrepancy either, with these two curves differing significantly. Indeed, the error bars shown in Fig. \ref{fig:timeevo}(b) represent the error of the mean over these classical trajectories, corresponding to a statistical significance of 99.5$\%$. For clarity, these error bars are only plotted for every ten points. This could be due to some remnants of integrability in the low-energy region that survive up to these excitation energies. However, that is not the case, as shown in Fig. \ref{fig:distributions}(b). Here we find that the population distributions of the coherent state and the average over 2000 classical trajectories coincide excellently, and yet the expectation values of observables do not.

These observations are even more striking in the case of collective systems such as the BH model, where the quantum-classical correspondence is usually established quite straightforwardly. In fact, quantum coherent state does not appreciably approach the symmetric equilibration value even when the number of particles, $N$, increases, as shown in Fig. \ref{fig:scaling_quantum}. In this figure we have represented the expectation value of $\hat{n}_{\max}$ in the quantum coherent state (red lines) for three different system sizes, $N=35,55$ and $75$ (from lighter to darker red). These expectation values reach a relatively stable value for $t\gtrsim 5000$, and this value is very far from the symmetric result $\overline{n}_{k}=0.25$. In fact, the deviations from $n_{k}=0.25$ generally increase with system size. We also note that the behavior of these curves as $N$ increases is not strictly monotonous. In the same figure we have also represented the value of $\hat{n}_{\max}(t)$ calculated for the classical average of trajectories discussed above, for the same values of $N$ (from lighter to darker green). The error bars correspond to the case $N=55$ and they represent three standard deviations from the mean. Error bars for other values of $N$ are qualitatively similar. This shows that with a likelihood of $99.5\%$ these two curves are not compatible, and the distance between them does not seem to decrease with $N$.

In other words, the striking difference between the quantum dynamics and that of an average of classical trajectories around $\langle \epsilon\rangle=-5.85$ shown in Fig. \ref{fig:timeevo}(b) cannot be attributed to differences in the LDOS. Following Eq. \eqref{eq:lta3}, these differences can only be due to the interplay between the eigenvalues $i_{n,r}$ and the population coefficients $c_{n,r}$. More precisely, such discrepancy is due to the fact that a generic coherent state significantly populates states with $i_{n,r}\neq 0$.

\section{Conclusions}\label{sec:conclusions}
We have presented an analysis of chaos, thermalization, and the quantum–classical correspondence in the collective four-site Bose–Hubbard model. Our results reveal a remarkably rich dynamical structure. By combining classical phase-space analysis, spectral diagnostics connected to quantum chaos, and real-time evolution of initial states of different nature, we have demonstrated that the system exhibits three distinct dynamical regimes: a low-energy symmetry-breaking phase with disconnected classical wells, an intermediate-energy region where ergodicity coexists with a seemingly persistent mismatch between quantum and classical long-time values, and a high-energy chaotic regime where correspondence is restored.

A key result of our work is the effective breakdown of the quantum–classical correspondence in the intermediate region above the first ESQPT. There, narrow classical phase-space bridges lead to intermittency and extremely slow equilibration, while quantum coherent states remain trapped in symmetry-breaking sectors due to the significant weight of imbalance-carrying eigenstates. This disagreement between classical and quantum dynamics is not attributable to differences in population distributions and, strikingly, does not diminish with increasing particle number: on the contrary, quantum deviations become more pronounced. These findings indicate that collective long-range models may exhibit strong finite-size effects that persist up to unexpectedly large system sizes. However, our results are restricted to the values of $N$ that we are able to numerically diagonalize, so we cannot completely discard the possibility that this mismatch disappears for macroscopic values of $N$. 

Overall, our work highlights that even in collective models, where classicality is often assumed to emerge straightforwardly, the route to the thermodynamic limit can be nontrivial. The interplay between ESQPT-induced phase-space structures, chaoticity, and symmetry-breaking leads to emergent dynamical phases that strongly impact equilibration and can challenge standard expectations of quantum–classical correspondence.

\acknowledgments
A.L.C., P.S., and P.C. acknowledge financial support from the Czech Science Foundation under project No. 25-16056S. A.L.C. also acknowledges support from the JUNIOR UK Fund project carried out at the Faculty of Mathematics and Physics, Charles University. A.R. acknowledges financial support by the Spanish grant
PID2022-136285NB-C31 funded by Ministerio de Ciencia e Innovación/Agencia Estatal de Investigación MCIN/AEI/10.13039/501100011033 and FEDER “A Way of Making Europe”. The data that support the findings of this article are openly available \cite{data}.

\end{document}